# Explaining the Performance of Collaborative Filtering Methods with Optimal Data Characteristics

Samin Poudel and Marwan Bikdash

Department of Computational Data Science and Engineering, North Carolina A & T University, Greensboro, USA

## ABSTRACT

*The performance of a Collaborative Filtering (CF) method is based on the properties of a User-Item Rating Matrix (URM). And the properties or Rating Data Characteristics (RDC) of a URM are constantly changing. Recent studies significantly explained the variation in the performances of CF methods resulted due to the change in URM using six or more RDC. Here, we found that the significant proportion of variation in the performances of different CF techniques can be accounted to two RDC only. The two RDC are the number of ratings per user or Information per User (IpU) and the number of ratings per item or Information per Item (IpI). And the performances of CF algorithms are quadratic to IpU (or IpI) for a square URM. The findings of this study are based on seven well-established CF methods and three popular public recommender datasets: 1M MovieLens, 25M MovieLens, and Yahoo! Music Rating datasets.*

## 1. INTRODUCTION

Collaborative Filtering (CF) recommendation system is widely used in domains like e-commerce, retail, social media, banking etc. CF methods predict the unknown ratings based on the known ratings data stored in a User-Item Rating Matrix (URM)[1]. Thus, the performance of a CF method relies on the Rating Data Characteristics (RDC) of the URM [2]–[4]. In other words, the URM properties can explain the variations in the performance of a CF method [5]–[7]. The RDC are mainly based on the structure of the URM, the rating frequency of the URM and the rating distribution in the URM [8], [9]. The RDC in the different recommendation domains can be very different [10]. Also, the RDC in a recommendation dataset are changing rapidly because of the exponential increase in the size of the data, resulting due to the development in the Internet and the web-based applications [11].

Many new CF methods are developed regularly with the pursuit of exploring new techniques that perform better than the existing ones. But there are very few works which study the variation in the performance of CF techniques based on RDC. In [12], the authors showed that the density impacts the performance of CF algorithms. Authors in [13], [14], and [15] showed that the accuracy of CF techniques improve with the increasing number of users, items and density.
In [8], authors studied the impact of six data characteristics in the performance of three CF algorithms in detail. They showed that the six data characteristics can significantly explain the variation in the performance of CF techniques for some datasets but not for all.

Authors in [5] studied the effect of six data characteristics on the robustness of three CF algorithms. Their study showed that around 80 percent variations in the effectiveness of shilling attacks on 3 CF models can be explained based on the six RDC. In [9], the combined effect of about ten RDC on the performances of CF based on the accuracy and the fairness were studied on several CF recommendation models. The study in [9] showed that the ten RDC explained variation in the performance in terms of accuracy better than the performance based on fairness.

From the available studies it is seen that, around six or more data characteristics are required to explain the significant proportion of variation in the performance of CF models resulted due to the change in the properties of a recommendation dataset. It is clear that the authors are being lured to use more and more data characteristics so that the variation in the performance of CF across different User-Item Rating Matrices (URMs) can be explained better. We do not find the studies that tend to find the optimal number of data characteristics capable of explaining the significant amount of variation in the performance of CF algorithms with the changing URM. Therefore, in this study we seek to find the minimum number of RDC that can explain the variation in the performance of seven CF models.

The study here is based on three well-known datasets (1M MovieLens [16],[17], 25M MovieLens [17], [18] and, the Yahoo! Music dataset [19] and seven well-established CF approaches :

1. Regularized Singular Value Decomposition denoted as SVD[12], [20] [Cacheda2011, Wilkinson1971]
2. SVD with bias terms denoted as SVD_b[21], [22]
3. Non-negative Matrix Factorization (NMF)[23]
4. Slope One [24]
5. Co-Clustering [25][George2005]
6. User based Nearest Neighbor (UNN)[5], [26]
7. Item based Nearest Neighbor (INN) [5], [26]

## 2. METHODOLOGY

Let, $R$ denotes the User-Item Rating Matrix (URM). Here, $m$ represents the size of the rows and hence the number of users, while $n$ represents the size of the columns, and hence the number of items.

Let $N_r$ denotes the total number of ratings or nonzero elements in a rating matrix. If $N_r$ is considered as the size of information in URM, then number of ratings per user can be defined as the Information per User (IpU). Mathematically,

$$\text{IpU} = \frac{N_r}{m} \quad \ldots \ldots \ldots (1)$$

Similarly, the number of ratings per item can be defined as the Information per Item (IpI). Mathematically,

$$\text{IpI} = \frac{N_r}{n} \quad \ldots \ldots \ldots (2)$$

A CF application process first divides a rating matrix R into training and test data, then learns a model from the training data and test the performance of the learned model using the test data. In this study, we implemented the CF algorithms using SURPRISE python library[21]. SURPRISE is an open-source python module for building and testing recommender systems with explicit rating data.

There are different predictive accuracy metrics, classification accuracy metrics, ranking accuracy metrics and others to evaluate the performance of a CF technique [27]–[29]. And, a universal RS does not exist which can perform better than other methods for all the evaluation metrics[30], [31]. However, the most popular and commonly used metric for prediction evaluation of a CF RS is predictive accuracy metric, Root Mean Squared Error (RMSE) [12], [32]. In this work, we have used the RMSE as the error measure of CF estimates. Let us define the performance measure of a CF method as the inverse of the error measure. Meaning,

$$\text{Performance } (P) = \frac{1}{\text{RMSE}} \quad \ldots \ldots \ldots (3)$$

Higher the RMSE of a CF method rating predictions lower is the performance $P$ and vice versa.

## 2.1. Overview of the Methodology

Here, we sample around 3000 User-Item Rating Matrices (URMs) randomly from each 1M MovieLens [16],[17], 25M MovieLens [17], [18] and, the Yahoo! Music dataset [19]. The number of rows (users) in the subsampled URMs ranged from 300 to 8000 users. Similarly, the number of columns (items) in the subsampled URMs ranged from 300 to 8000 items. The applied random sampling procedure is similar to the one mentioned in [8]. The sampled rating matrices vary widely in most of the URM characteristics mentioned in [Deldjoo2021]. Then, we apply seven CF methods to each of the sampled URM and compute the performance measure based on Eq. (3). After that, we plot for the Performance of CF methods versus Information per User (IpU) at constant Information per Item (IpI). Similarly, we plot for the Performance of CF methods versus IpI at constant IpU. Next, we perform the regression analysis of CF Performance versus IpU and IpI, and then discuss the results.

The URMs extracted from 1M MovieLens have rating data in numerical discrete scale from 1.0 to 5.0 in steps of 1.0. The URMs extracted from 25M MovieLens have rating data in numerical discrete scale from 0.5 to 5.0 in steps of 0.5. And the ones extracted from Yahoo! Music rating data have discrete data scale of 1.0 to 100.0.

## 3. Plots of CF performance Versus Rating Data Characteristics

During the analysis of the CF Performance versus different rating data characteristics, we found that the CF Performance is linear to log (IpI) at constant IpU as shown in Figure 1. Similarly, the CF Performance is linear to log (IpU) at constant IpI as shown in Figure 2. Therefore, we can define a CF performance regression model as:

$$P = a_0 + a_1 \log(\text{IpU}) + a_2 \log(\text{IpI}) + a_3 \log(\text{IpU}) \log(\text{IpI}) \quad \ldots \ldots \ldots (4)$$

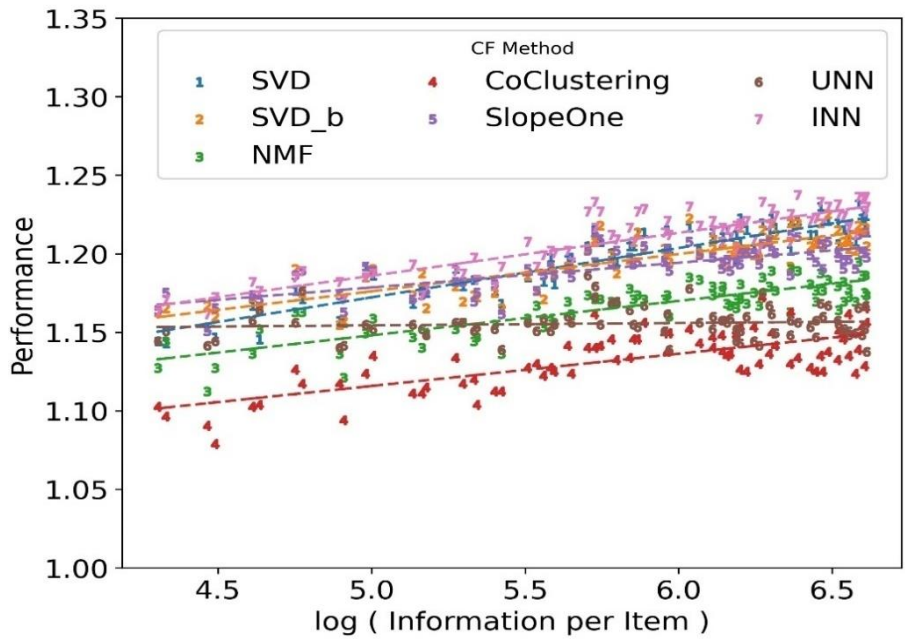

Figure 1. CF performance versus log(IpI) at constant IpU of 58.55

The plots in Figures 1 and 2 give an insight that the two RDC, Information per User and Information per Item have capability of explaining the significant proportion of the variation in the performance of all seven CF methods. To prove this insight, we need to perform the regression analysis based on Eq. (4), which is to follow in next section.

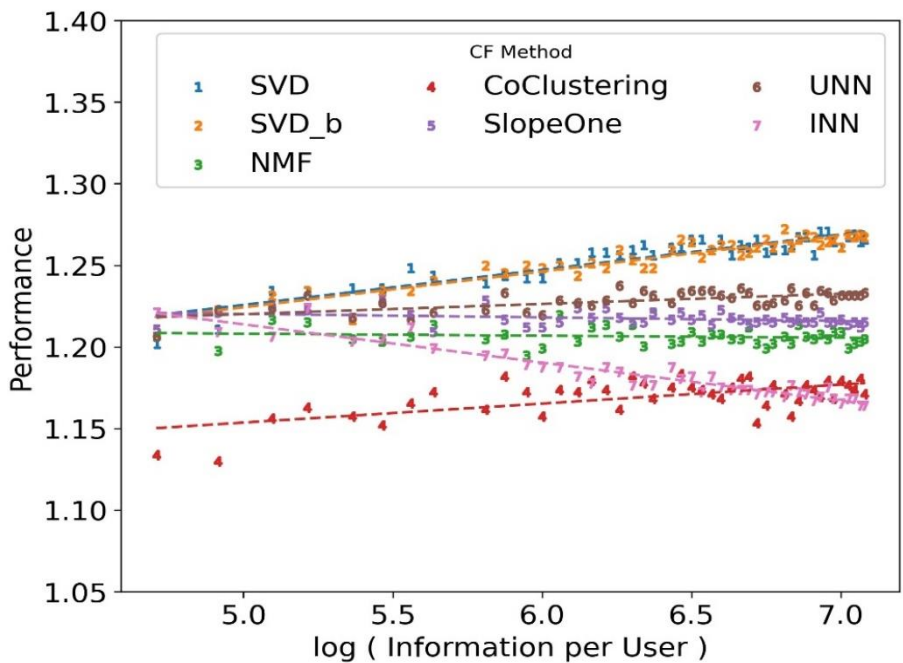

Figure 2. CF performance versus log(IpU) at constant IpI of 55.55

## 3.1. Regression Analysis of CF Performance versus Rating Data Characteristics

In this section, we perform the regression analysis of CF performance based on Eq. (4). The results of the regression analysis along with the evaluation measures for explaining the variation, adjusted $R^2$ are tabulated in Tables 1, 2, and 3. The values of adjusted $R^2$ can be 0 in minimum and 1 in maximum. The higher the value of adjusted $R^2$, the higher the capability of the independent variables to explain the variation in the dependent variable of a regression model. All the coefficients and constant reported in Tables 1, 2, and3 had p-values less than 0.01 which makes the regression analysis results statistically significant.

Here, The Table 1 shows the results of regression analysis using Eq. (4) based on the sampled URMs from 25M MovieLens data. The results show that the IpU and IpI have significant explanatory power (70 to 95 percent) to explain the variance in the performance of all CF models except for Slope-One and NMF when considering the URMs from 25M MovieLens data.

Table 1. Regression Analysis based on Eq. (4) using 25M MovieLens data.

| CF Method | Coefficient of log (IpU) | Coefficient of log (IPI) | Coefficient of log (IpU) * log (IPI) | constant | $R^2$ (Adjusted) |
|---|---|---|---|---|---|
| UNN | 0.0121 | -0.0017 | 0.0019 | 0.9843 | 0.82 |
| INN | -0.0902 | -0.0318 | 0.0151 | 1.33 | 0.89 |
| SVD | 0.0283 | 0.118 | 0.0040 | 0.9205 | 0.94 |
| SVD_b | 0.0149 | -0.0024 | 0.0067 | 0.9891 | 0.95 |
| CoClustering | 0.0452 | 0.0361 | -0.0044 | 0.8561 | 0.71 |
| Slope-One | 0.0390 | 0.0310 | -0.0045 | 0.9577 | 0.51 |
| NMF | 0.0558 | 0.0442 | -0.0066 | 0.8461 | 0.57 |

The Table 2 shows the results of regression analysis using model in Eq (4) based on the sampled URMs from 1M MovieLens data. The results show that the IpU and IpI have significant explanatory power (70 to 92 percent) to explain the variance in the performance of all seven CF models when considering the URMs from 1M MovieLens data.

Table 2. Regression Analysis based on Eq. (4) using 1M MovieLens data.

| CF Method | Coefficient of log (IpU) | Coefficient of log (IPI) | Coefficient of log (IpU) * log (IPI) | constant | $R^2$ (Adjusted) |
|---|---|---|---|---|---|
| UNN | 0.0341 | 0.0245 | -0.0028 | 0.8153 | 0.82 |
| INN | -0.0195 | 0.0072 | 0.0057 | 0.8961 | 0.84 |
| SVD | 0.1592 | 0.1288 | -0.0199 | 0.1831 | 0.92 |
| SVD_b | -0.0166 | -0.0195 | 0.0088 | 1.0808 | 0.90 |
| CoClustering | 0.0196 | 0.0056 | 0.0031 | 0.8889 | 0.87 |
| Slope-One | 0.0398 | 0.0283 | -0.0038 | 0.8631 | 0.74 |
| NMF | 0.0584 | 0.0376 | -0.0042 | 0.7179 | 0.88 |

The Table 3 shows the results of regression analysis using model in (4) based on the sampled URMs from Yahoo! Music Rating data. The results show that the IpU and IpI have significant explanatory power (70 to 93) to explain the variance in the performance of all CF models except SVD_b when considering the URMs from Yahoo Music Rating Data.

Table 3. Regression Analysis based on Eq. (4) using Yahoo Music Rating data.

| CF Method | Coefficient of log (IpU) | Coefficient of log (IPI) | Coefficient of log (IpU) * log (IPI) | constant | $R^2$ (Adjusted) |
|---|---|---|---|---|---|
| UNN | 0.0006 | 0.0007 | 0 | 0.0316 | 0.90 |
| INN | -0.0034 | -0.0023 | 0.008 | 0.0523 | 0.93 |
| SVD | 0.0023 | -0.0022 | -0.0006 | 0.0355 | 0.71 |
| SVD_b | 0.0015 | 0.0015 | -0.004 | 0.0336 | 0.35 |
| CoClustering | 0.0016 | 0.0018 | -0.0002 | 0.0303 | 0.75 |
| Slope-One | 0.0010 | 0.0011 | -0.0001 | 0.0347 | 0.70 |
| NMF | 0.0004 | 0.0004 | -0.0001 | 0.0127 | 0.79 |

From the regression analysis results, it is clear that the variation in the performance of different CF models resulted due to changes in URM can be explained significantly with the IpU and IpI only. The results in this study involving two RDC is comparable to the study in [8] using six RDC for memory-based CF methods (UNN and INN). And the results much better for matrix factorization based methods (SVD, NMF) compared to the result in[8].

## 3.2. Discussion of CF Performance versus Rating Data Characteristics

The regression analysis and the evaluation presented in the Section 3.1 suggests Eq. (4) as a statistically significant model to explain the variation in the performance of CF. Eq. (4) is re-stated as below:

$$P = a_0 + a_1 \log(\text{IpU}) + a_2 \log(\text{IpI}) + a_3 \log(\text{IpU}) \log(\text{IpI}) \quad \dots\dots\dots(5)$$

The model in Eq. (5) has significant power to explain the variability in the performance of CF methods based on two RDC only as opposed to the work found in the literature, which uses six or more RDC[5], [8], [9].

The partial derivative of the Eq. (5) with IpU gives:

$$\frac{\delta P}{\delta \text{IpU}} = \frac{1}{\text{IpU}}(a_1 + a_3 \log(\text{IpI})) = \frac{c_1}{\text{IpU}} \quad \dots\dots\dots(6)$$

which shows the rate of change in CF performance $P$ per IpU is inversely proportional to the information per user IpU at constant IpI. Similarly, the partial derivative of the Eq. (5) with IpI gives:

$$\frac{\delta P}{\delta \text{IpI}} = \frac{1}{\text{IpI}}(a_2 + a_3 \log(\text{IpU})) = \frac{c_2}{\text{IpI}} \quad \dots\dots\dots(7)$$

which shows the rate of change in CF performance $P$ per IpI is inversely proportional to the information per user IpI at constant IpU.

For a square URM, IpU = IpI. Then, the model in (5) can be written as

$$\begin{cases} P = a_0 + (a_1 + a_2) \log(\text{IpU}) + a_3 \big(\log(\text{IpU})\big)^2 \\ = a_0 + (a_1 + a_2) \log(\text{IpI}) + a_3 \big(\log(\text{IpI})\big)^2 \end{cases} \quad \dots\dots\dots(8)$$

Hence, from the model in (8), we can state that the performance of a CF algorithm is quadratic in IpU or IpI for a square URM.

# 4. CONCLUSION AND FUTURE WORK

In this study, we checked the capability of two Rating Data Characteristics (RDC) to explain the variation in the performance of CF methods. The two RDC are the number of ratings per user or Information per User (IpU) and the number of ratings per item or Information per Item (IpI). Our study showed that the performance of seven CF methods considered in this study varied linearly with log (IpU) at constant IpI. Similarly, the performance of all CF methods varied linearly with log (IpI) at constant IpU. The extensive simulation and regression analysis disclosed that the significant proportion of the variations in the performance of seven CF methods resulted due to change in URMs can be explained with the IpU and IpI only. The results in this study are based on the 1M MovieLens Dataset, the 25M MovieLens Dataset and, the Yahoo Music Rating Dataset.

Our study here is based on seven well established CF methods and three public recommendation datasets. The results are not tested for machine learning algorithms beyond CF recommendations. Also, the conclusions of this study are empirical, the theoretical foundation behind the results still needs to be explored. And the acceptability of the results for User-Item Rating Matrices from different recommender domains is to be tested.

## AUTHORS

**Samin Poudel** received PhD in Computational Data Science and Engineering from NorthCarolina A&T State University in 2022. He did his undergraduate in Nepal. His research interests include but not limited to data analytics, data mining, machine learning, developing models, and optimizing techniques based on data.

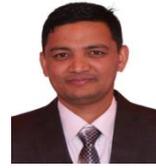

**Marwan Bikdash** received the MEng and PhD degrees in electrical engineering from Virginia Tech in 1990 and 1993, respectively.He is currently a professor and the chair of the Department of Computational DataScience and Engineering, North Carolina A&T State University. He teaches and conducts research in signals and systems, computational intelligence, and modelling and simulations of systems with applications in health, energy, and engineering. He has authored over 150 journal and conference papers. He has supported, advised, and graduated over 50 master and PhD students. His projects have been funded by the Jet Propulsion Laboratory, Defense Threat Reduction Agency, Army ResearchLab, NASA, National Science Foundation, the Office of Naval Research, Boeing Inc., Hewlett Packard, National Renewable Energy Laboratories, the Army Construction Engineering Research Laboratory, and others.

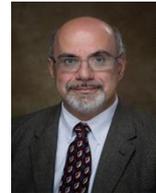